\documentclass{article}
\usepackage{hyperref}
\usepackage{graphicx}
\usepackage{dcolumn}
\usepackage{bm}
\usepackage{color}
\usepackage{amsmath}
\usepackage[utf8]{inputenc}
\usepackage[T1]{fontenc}
\usepackage{epstopdf}
\usepackage{multirow, multicol,times,subfigure, setspace, rotating, url}
\usepackage{amssymb}
\usepackage{pifont}
\usepackage{enumerate}
\usepackage[superscript, biblabel, nomove]{cite}
\begin{document}
\title{Measuring and utilizing temporal network dissimilarity}
\author{\normalsize
 Xiu-Xiu Zhan{\small$^{\mbox{1}}$},
 Chuang Liu{\small$^{\mbox{1}}$},
 Zhipeng Wang{\small$^{\mbox{1}}$},
 Huijuan Wang{\small$^{\mbox{2}}$},
 Petter Holme{\small$^{\mbox{3}}$},
 Zi-Ke Zhang{\small$^{\mbox{4,1}}$}
}

\maketitle

\vspace{-5mm}

\noindent
$^1$ Research Center for Complexity Sciences, Hangzhou Normal University, Hangzhou 311121, PR China \\
$^2$ Faculty of Electrical Engineering, Mathematics, and Computer Science, Delft University of Technology, Mekelweg 4, 2628 CD, Delft, The Netherlands\\
$^3$ Tokyo Tech World Hub Research Initiative, Institute of Innovative Research, Tokyo Institute of Technology, Tokyo 226-8503, Japan\\
$^4$ College of Media and International Culture, Zhejiang University, Hangzhou 310058, PR China\\



\begin{abstract}
Quantifying the structural and functional differences of temporal networks is a fundamental and challenging problem in the era of big data. This work proposes a temporal dissimilarity measure for temporal network comparison based on the fastest arrival distance distribution and spectral entropy based Jensen-Shannon divergence. Experimental results on both synthetic and empirical temporal networks show that the proposed measure could discriminate diverse temporal networks with different structures by capturing various topological and temporal properties. Moreover, the proposed measure can discern the functional distinctions and is found effective applications in temporal network classification and spreadability discrimination.
\end{abstract}

%
%
\thispagestyle{empty}

\section*{Introduction}

Complex networks are the leading framework for understanding real-world complex systems, ranging from physical and nervous systems~\cite{barabasi2013network}, biological and chemical reactions~\cite{newman2018networks,liu2020computational}, to financial~\cite{caldarelli2013reconstructing} and social platforms~\cite{li2017fundamental,li2017simple}. The primary entities of a complex network are represented by nodes, whereas links indicate relationships or interactions among nodes. Despite the substantial advances in expressing and analyzing entities with simplex~\cite{battiston2020networks}, multiple~\cite{boccaletti2014structure, de2016physics} or signed~\cite{leskovec2010signed, facchetti2011computing} relationships across various systems, the temporal networks, which characterize the evolutionary process of dynamic systems~\cite{masuda2017temporal,holme2019temporal,li2020evolution}, call for an immediate solution to take into account the precise record when each interaction happens or perishes and analyze the underlying mechanisms that drive the emergence of such structural differences~\cite{tang2020predictability}.


Network comparison~\cite{mellor2019graph, pierri2020topology}---aiming to give a principled answer to the simple question of how different two networks are---is strikingly challenging for temporal networks~\cite{tantardini2019comparing}. One approach is to aggregate a temporal network into a static network (Figure~\ref{fig:TN-toy}A-\ref{fig:TN-toy}C), and adopt associated static network comparison methodologies ~\cite{bagrow2019information,martinez2019comparing}. However, this simplification ignores the time ordering of contacts which may capsule rich and meaningful information, such as the daily and circadian rhythms of interactions~\cite{masuda2017temporal,holme2019temporal}. As a consequence, the corresponding comparison methods \cite{hand2007principles,koutra2013deltacon,malod2015graal,de2016spectral,schieber2017quantification,gera2018identifying,tantardini2019comparing} that are tailored to static networks, therefore, cannot fully capture the dynamical patterns. For example, as shown in Figure~\ref{fig:TN-toy}, two distinctive temporal networks $G_1$ (Figure~\ref{fig:TN-toy}A) and $G_2$ (Figure~\ref{fig:TN-toy}B) have the same static network structure (Figure~\ref{fig:TN-toy}C). The differences between $G_1$ and $G_2$ will be zero according to static network-based comparison methods. Therefore, to address this problem, it is essential to take into account temporal information to discern slight structural differences and crucial underlying dynamic patterns.


In this work, we explore the fastest arrival distance (FAD) and spectral entropy based Jensen-Shannon divergence~\cite{wu2014path, de2016spectral, schieber2017quantification} to characterize the dissimilarity of temporal networks (Figure~\ref{fig:TN-toy}D-\ref{fig:TN-toy}G). Furthermore, we apply the derived measure to network classification and spreading dynamics to demonstrate its validity. To perform the analysis, we use both synthetic and real-world datasets to evaluate the effectiveness of the proposed measure in discriminating the differences between temporal networks. These detailed datasets allow us to examine the impact of time-varying interactions on diverse temporal paradigms with different network sizes and time scales.


%



\section*{Results}\label{Results}
\subsection*{Temporal Dissimilarity Characterization}
\begin{figure*}[!ht]
\centering
	\includegraphics[width=18cm]{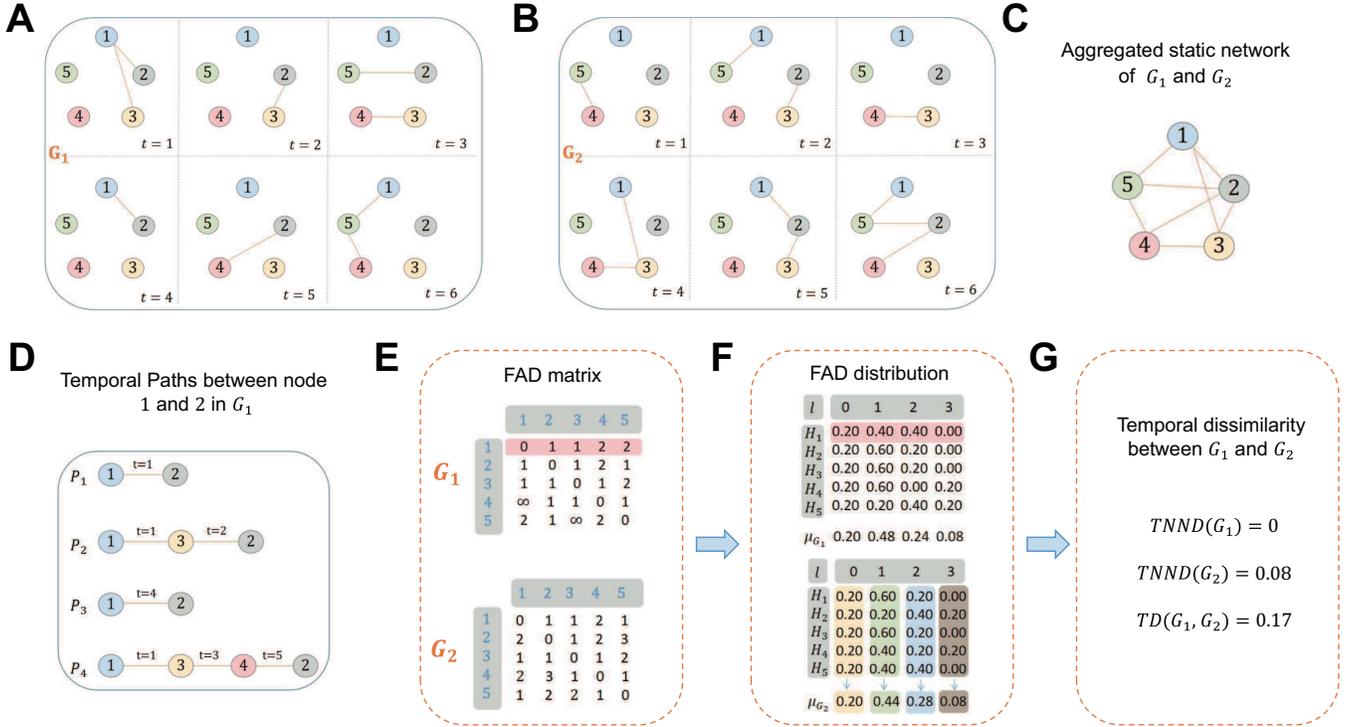}
 \caption{\textbf{Illustration of temporal dissimilarity characterization}. (A, B) Visualization of two temporal networks $G_1$ and $G_2$, which share the same number of nodes $N$ and window size $T$;
 (C) The aggregated static network of $G_1$ and $G_2$; (D) Temporal paths between nodes 1  and 2 in $G_1$. (E-G) Illustration of computing the dissimilarity between $G_1$ and $G_2$. Firstly, we compute the FAD matrix, i.e., the fastest arrival distance between node pairs, for $G_1$ and $G_2$, respectively. Here, we denote $l_{\max}(G)$ to be the maximal FAD among all fastest arrival paths in temporal network $G$. When a fastest arrive path does not exist between two nodes thus the two nodes are not reachable, the corresponding FAD is set as $l_{\max}(G)+1$. This happens in $G_1$, and $l_{\max}(G_1)=2$. Secondly, the FAD distribution $H_i$ of node $i$ and $\mu_{G_1} (\mu_{G_2})$ of the whole network are computed according to the FAD matrix. Finally, the temporal node dispersion (TNND) and temporal dissimilarity (${\rm TD}(G_1, G_2)$) are given based on Eqs.~\eqref{TNND} and \eqref{TD-EQ}.}
 \label{fig:TN-toy}
\end{figure*}

For a given temporal network $G=(V,E)$ in a time window $[0, T]$, $V$ represents the node set, and $E=\{(i, j, t), t\in[0, T], i, j\in V \}$ is the contact set. An element $(i, j, t)$ indicates that there is a contact between node $i$ and $j$ at time step $t$. The number of nodes and temporal edges is given by $N$  and $C$, respectively. The aggregated static network $G^s$ of a temporal network $G$ is the graph with the same node set and an edge between all pairs of nodes that have at least one temporal edge in $E$.  The number of edges in the aggregated $G^s$ is denoted by $M$.  In Figure~\ref{fig:TN-toy}A-\ref{fig:TN-toy}C, we show examples of two temporal networks with five nodes and six timestamps, i.e., $G_1$ and $G_2$, and the aggregated static network.

A temporal path $P$ in $G$ is a node sequence $P=\langle i_\nu \rangle_{\nu=1}^{n+1}$, where $(i_\nu, i_{\nu+1}, t)\in E$ for all $\nu\in[1,n]$. The length of $P$, $l(P)$, is defined as the number of contacts in the path. In this work, we shall base our temporal dissimilarity on the concept of \textit{fastest arrival paths}, which can capture both topological and temporal proximity between nodes and has been proved to be an effective way for temporal network topology mining~\cite{wu2014path}. The fastest arrival path between nodes $i_s$ and $i_t$ is the temporal path from $i_s$ starting at $t=0$ that reaches $i_t$ the earliest. Meanwhile, we adopt the fastest arrival distance (FAD) as the length of the fastest arrival path between two nodes. The temporal paths between nodes 1 and 2 in $G_1$ are shown in Figure~\ref{fig:TN-toy}D, i.e., $P_1$, \dots, $P_4$. According to the definition of fastest arrival path, we can obtain that $P_1$ is the fastest arrival path between nodes 1 and 2, and thus the FAD value between them is $l(P_1)=1$.

We then define the FAD distribution of node $i$ as $H_i=\{h_{i}(q)\}$, where $h_{i}(q)$ is the fraction of nodes with FAD $q$ from $i$. We use $l_{\max}$ to denote the maximal FAD among all fastest arrival paths in a temporal network. If every node pair is reachable through a fastest arrival path in a temporal network, the dimension of $H_i$ is $1 \times l_{\max}$. If there are node pairs in a network that are not reachable, we define the maximal FAD between reachable node pairs as $l_{\max}$ and the infinite FAD is defined as $l_{\max}+1$. In this case, the dimension of $H_i$ is $1 \times (l_{\max}+1)$. 
The node specific FAD distribution $H_i$ contains the connectivity heterogeneity of node $i$.
Subsequently, we adopt Jensen-Shannon divergence to characterize the connectivity heterogeneity of a temporal network based on FAD. The temporal network node dispersion (TNND) reads:
\begin{equation}\label{TNND}
\centering
{\rm TNND}(G)=\frac{J(H_1, H_2, \cdots, H_N)}{\log(l_{\max} + 1)},
\end{equation}
where $J(H_1, H_2, \cdots, H_N)=\frac{1}{N}\sum_{i,q}h_{i}(q)\log(h_{i}(q)/u_q)$ is the Jensen-Shannon divergence. The term $\mu_q=\frac{1}{N}\sum_{i=1}^{N}h_{i}(q)$ is the average over $\{h_{1}(q), h_{2}(q), \cdots, h_{N}(q)\}$. It represents the fraction of source destination node pairs with FAD $q$. The average FAD distribution of a network, i.e., the probability distribution of the FAD of a random source destination pair, is given by $\mu=\{\mu_1, \mu_2,\cdots, \mu_k, \cdots\}$. The dimension of $\mu$ is either $1 \times l_{\max}$ or $1 \times (l_{\max}+1)$. Generally, a larger TNND implies higher connection diversity among the nodes.

Finally, for two given networks $G_1$ and $G_2$, their temporal dissimilarity, ${\rm TD}(G_1, G_2)$, reads
\begin{equation}\label{TD-EQ}
\centering
{\rm TD}(G_1, G_2)=\omega_1\sqrt{\frac{J(\mu_{G_1}, \mu_{G_2})}{\log2}} + \omega_2|\sqrt{{\rm TNND}(G_1)}-\sqrt{{\rm TNND}(G_2)}|,
\end{equation}
where ${\rm TNND}(G_1)$ and ${\rm TNND}(G_2)$ represent the temporal network node dispersion, and $\mu_{G_1}$ and $\mu_{G_2}$ are the average FAD distribution of $G_1$ and $G_2$, respectively. $\omega_1$, $\omega_2 \in [0, 1]$ (satisfying $\omega_1 +\omega_2=1$) are tunable parameters to measure the contribution of the average FAD distribution and TNND difference, respectively. For the sake of simplification, we use $\omega_1=\omega_2=0.5$ in the following analyses. Generally, TD falls in $[0,1]$ and the larger TD is, the more dissimilar the networks will be. For the extreme cases, ${\rm TD}=1$ suggests the two networks are completely different with each other, and vice verse for ${\rm TD}=0$. The detailed procedure of computing the dissimilarity between two exampled temporal networks $G_1$ (Figure~\ref{fig:TN-toy}A) and $G_2$ (Figure~\ref{fig:TN-toy}B) is shown in Figure~\ref{fig:TN-toy}E-\ref{fig:TN-toy}F.
It is found that ${\rm TD}(G_1, G_2)=0.17$, indicating that the two temporal networks are actually quite dissimilar as there are only five nodes. It is worth noting that, the dissimilarity value between them is zero based on static network comparison methods, because they share the same aggregated static network structure (Figure~\ref{fig:TN-toy}C). Therefore, it is of vital importance and urgent necessity to consider network temporality when conducting the comparisons.

\subsection*{Comparisons on Synthetic temporal networks }
To verify the validity of temporal dissimilarity (TD) metric, we perform comparisons on temporal synthetic networks. The objective is to explore whether the temporal dissimilarity could distinguish synthetic networks with different parameters/properties. The activity driven model~\cite{perra2012activity} (see \textbf{Materials and Methods}) is used to generate a collection of temporal networks $ G (F(a), m)=\{ G ^{t}\}_{t=0}^{T}$ via the tunable function $F(a)$ and parameter $m$. The function $F(a)$ controls the node activity distribution, and $m$ determines the number of contacts that every active node releases at each discrete time step $t$. Larger $m$ shall result in a temporal network with more contacts. The distribution of $F(a)$ determines the heterogeneity of activation rates of nodes. In this work, we select two representative types of activity distribution, i.e., (i) the uniform distribution $F(a) = 1/a_{\max}$, for $0\leq a \leq a_{\max} \leq 1,$
where $a$ is the active rate of a node and here we choose to use $a_{\max}=1$, and (ii) the power-law distribution $F(a) = (r-1)a^{-r}$, where $r > 2, 0 < a < 1$.

\begin{figure*}[!ht]
\centering
	\includegraphics[width=0.8\textwidth]{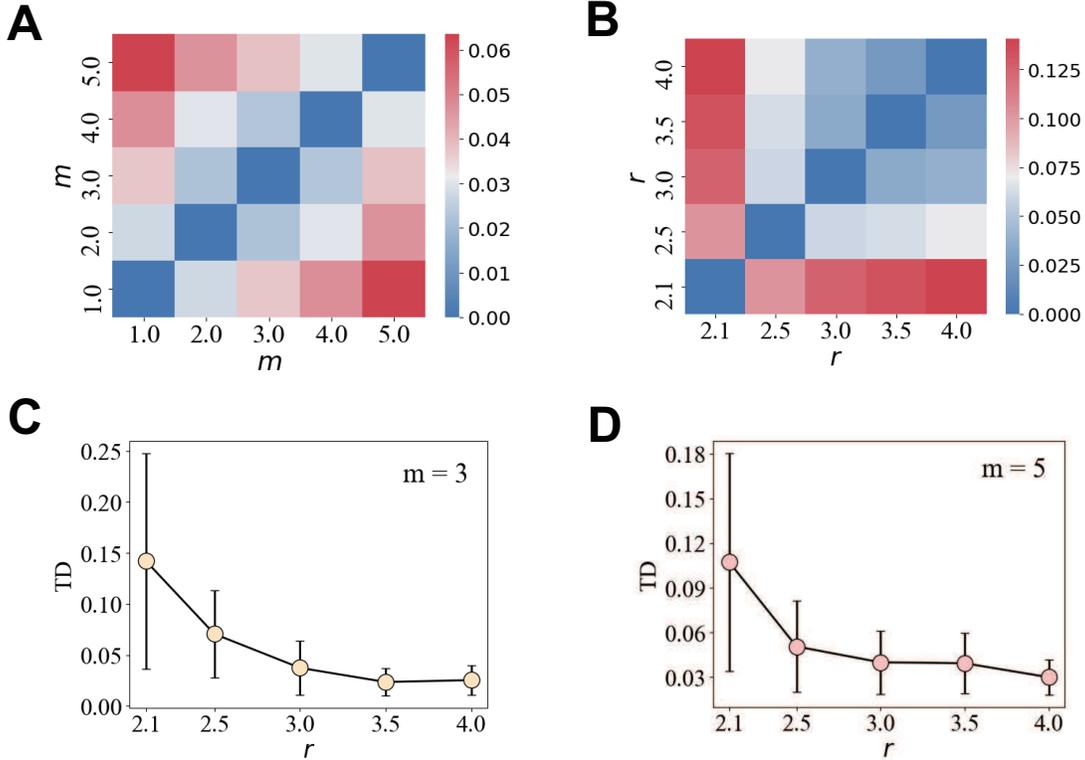}
 \caption{\textbf{Comparisons on synthetic temporal networks}. (A) Temporal dissimilarity between synthetic networks generated by the activity driven model with node activity following the uniform distribution and different $m$. (B) Temporal dissimilarity TD between synthetic networks generated by the activity driven model with node activity following the power-law distribution and different $r$. Here, we set $m=3$. (C, D) Temporal dissimilarity TD between networks generated by the uniform activity distribution and those generated by a power-law activity distribution as a function of $r$ when $m$ is set as 3 and 5 for (C) and (D), respectively. Here, each dissimilarity value in every sub-figure is the averaged over $100$ realizations. The network size and time scale are set as $N=300$, $T=30000$, respectively.}
 \label{fig:uni1-pl1}
\end{figure*}

Figure~\ref{fig:uni1-pl1}A shows the comparison results on synthetic temporal networks generated by the uniform activity distribution for different values of $m$. It indicates that the temporal dissimilarity tends to be small for the networks generated with similar $m$, suggesting that networks generated with similar $m$ would be more akin to each other and vice versa. 
Comparatively, Figure~\ref{fig:uni1-pl1}B shows the temporal dissimilarities of temporal networks generated by different heterogeneity parameters $r$. It shows that the temporal dissimilarity between temporal networks with closer $r$ are inclined to be smaller, suggesting that networks with closer $r$ tend to be similar. Here, the network size $N$ is $300$, and the time window size $T$ is $30000$. We can also see similar patterns for different network sizes, time window sizes, as well as other parameters (see Figures S1-S6 in \textbf{Supplementary Note 1}).

In addition to the analysis in the main text, we compare the temporal dissimilarities between networks derived from uniform and power-law distributions in Figures~\ref{fig:uni1-pl1}C-\ref{fig:uni1-pl1}D and also Figure~S7 in \textbf{Supplementary Note 1}. It shows that networks generated by the power-law activity distribution with a higher $r$ are more similar to those generated by the uniform activity distribution. This is
because a higher $r$ can result in networks with more homogeneous activity distribution, hence are more similar to the networks generated by uniform activity distribution. These observations in synthetic networks suggest TD to be a powerful index to discriminate the synthetic temporal networks.

\subsection*{Comparisons on real-world temporal networks}
Here, we introduce 17 representative temporal networks~\cite{paranjape2017motifs,michalski2011matching,van2012making,fournet2014contact,mastrandrea2015contact,stehle2011high,isella2011s,genois2018can}, including five email networks and 12 physical contact networks, to further validate the effectiveness of the proposed method (for details of datasets (Table 1), see \textbf{Materials and Methods}). We begin by comparing a temporal network with its aggregated static network. Given a static network, different from the temporal network, we alternatively use the shortest path distance distribution ($H_i^s=\{h_{i}^s(q)\}$, where $h_{i}^s(q)$ is denoted as the fraction of nodes that connect to $i$ at shortest path distance $q$ instead of FAD distribution.
We then substitute the FAD distribution ($H_i$) by shortest path distance distribution ($H_i^s$) in Eq.~\eqref{TNND} for the aggregated static network, and then update Eq.~\eqref{TD-EQ} accordingly (see \textbf{Supplementary Note 2} for details). 

\begin{figure*}[!ht]
\centering
	\includegraphics[width=18.5cm]{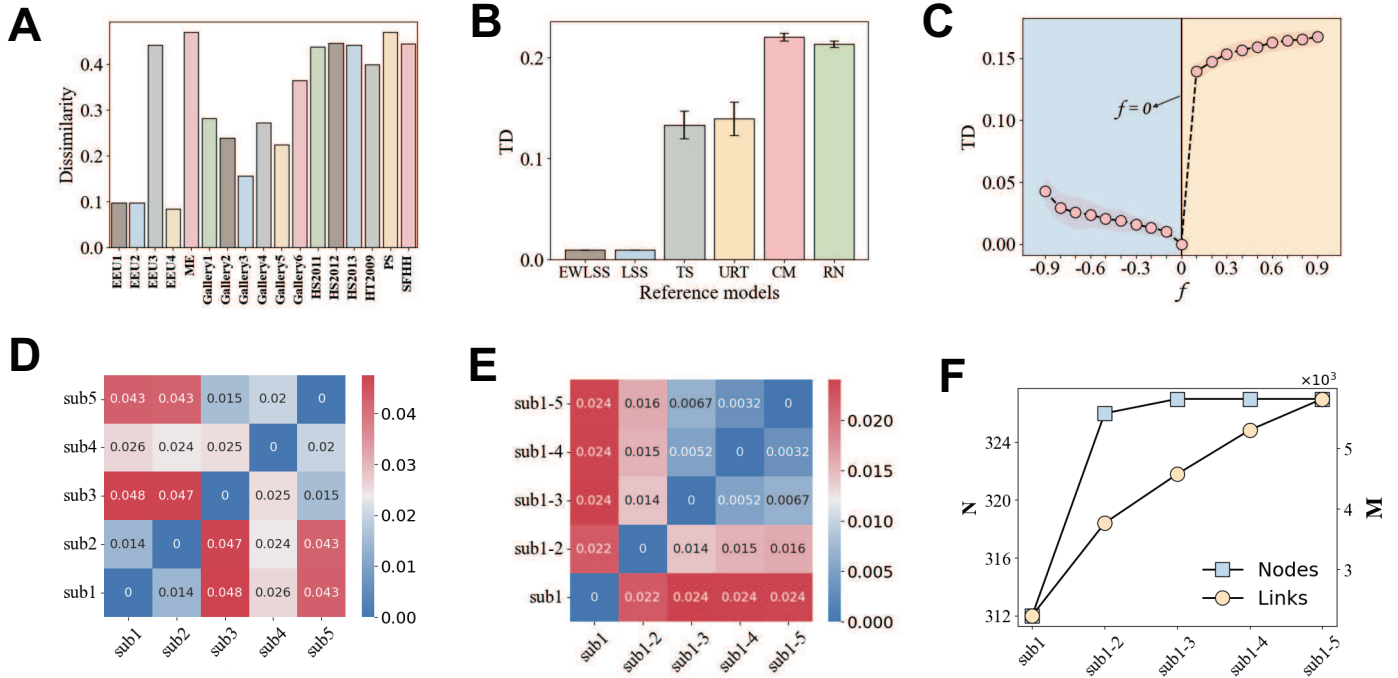}
 \caption{\textbf{Comparisons on real temporal networks} . (A) Dissimilarity between a temporal network and its aggregated static network; (B) Temporal dissimilarity between the original data and the temporal networks derived from reference models. (C) Temporal dissimilarity between the original and perturbed temporal networks. $f$ denotes the fraction of contacts added ($f>0$) or deleted ($f<0$) from the original temporal network ($f=0$). Each data point is averaged over 100 realizations. (D-E) Temporal dissimilarity between different daily divided sub-networks, e.g., \textit{sub1} and \textit{sub2} represent the temporal sub-networks of the first and second day, respectively. \textit{sub1-2} represents the cumulative temporal sub-network for the first two days. (F) The number of nodes ($N$) and links ($M$) of different sub-networks.
 (B-F) show comparisons on \textit{HS2013}, for more results of other networks, see Figure~S8-S10 in \textbf{Supplementary Note 2}. }
 \label{fig:static-tempo}
\end{figure*}

The resulting dissimilarity between an empirical temporal network and its aggregated static network is shown in Figure~\ref{fig:static-tempo}A. It is found that all dissimilarities are larger than zero, most of which are higher than $0.4$. It unveils that, in general, there is significant difference between the temporal network and its aggregated static network. Therefore, one might miss to obtain appropriate understanding and regularities by using static network to address time-varying interactions if the rich temporal information is ignored.

\subsubsection*{Comparisons based on temporal reference models}
We compare each of the empirical temporal networks with their reference models~\cite{kivela2012multiscale}, which are obtained by specific reshuffled methods (see \textbf{Materials and Methods}).
For each reference model, topological and temporal correlations are in various degrees of destruction (Table~\ref{refmodel}). For \textit{EWLSS}, \textit{TS} and \textit{URT}, all topological correlations are preserved, while the temporal correlations are destroyed to different extents. For \textit{LSS}, \textit{CM} and \textit{RN}, the temporal and topological correlations are both eliminated in varying degrees. Take \textit{HS2013} as an example, Figure~\ref{fig:static-tempo}B shows that the temporal dissimilarities between the original temporal network and the reference models are very small for \textit{EWLSS} and \textit{LSS}, while the dissimilarities are much larger for \textit{TS}, \textit{URT}, \textit{CM} and \textit{RN}. It can be found that, although \textit{TS} only eliminates the same number of properties as \textit{LSS}, it shall lead to higher dissimilarity as it destroys the temporal information (sorted sequence
list on each link) other than static topology correlation (weight). Comparatively, even though \textit{RN} discards the most number of properties, the resulting network does not show significantly different from that derived by \textit{CM}, suggesting that degree-degree correlation may not be a primary factor in a time-varying environment. Similar patterns can also be found for other datasets in Figure~S8 in \textbf{Supplementary Note 2}. Therefore, it can be concluded that the more temporal properties are preserved, the more similar it will be with the original temporal network.

\subsubsection*{Comparisons based on network perturbation and evolution}

We then assess the discriminative ability of the proposed measure by network perturbation method. That is to say, For a given temporal network, we randomly add or delete a fraction of contacts ($f$) and then compare the differences. Figure~\ref{fig:static-tempo}C shows the temporal dissimilarity between the original data and gradually perturbed network for \textit{HS2013}. The negative $f$ corresponds to the deletion process, and vice verse. Large $f$ indicates that more contacts are added ($f>0$) or deleted ($f<0$), resulting in
larger differences with respect to the original network ($f=0$). Note that, for networks with many contacts (large $C/N$ in Table~\ref{TB:1}), 	the disparity will even be reduced rather than increased, when adding new contacts (see Figure S9 in \textbf{Supplementary Note 2}), suggesting that the ceiling effect~\cite{rifkin2005ceiling} of time-varying interactions governs the structural differences of temporal networks.

Furthermore, we examine the temporal dissimilarity from the perspective of network evolution.
We divide a temporal network into sub-networks according to the time order of contacts. Take \textit{HS2013} for instance, we separate it into five distinct sub-networks where each contains all the contacts in one single day as the observation window is five days. We then use \textit{sub1} and \textit{sub2} to respectively represent the temporal sub-networks of the first and second day, and \textit{sub1-2} to represent the temporal sub-network of the first two days. Similar symbols are also described for other sub-networks. Figure~\ref{fig:static-tempo}D shows that, in general, sub-networks that are closer in time are more similar to each other, e.g., the temporal dissimilarity between \textit{sub1} and \textit{sub2} (${\rm TD}_{\textit{sub1},\textit{sub2}}$) is much smaller than those between \textit{sub1} and others. This is further enhanced in Figure~\ref{fig:static-tempo}E that a sub-network that forms earlier is quite different from the whole temporal network (\textit{sub1-5}) which cumulatively considers all the contacts. It is worth noting that, the short-term temporal dissimilarities (${\rm TD}_{\textit{sub1},\textit{sub3}}=0.048$ and ${\rm TD}_{\textit{sub2},\textit{sub3}}=0.047$) are unexceptional large, suggesting that the temporal interactions have experienced tremendous changes in the short run. Figure~\ref{fig:static-tempo}F shows that, although the number of nodes ($N$) are the same in \textit{sub1-4} and \textit{sub1-5}, the number of links ($M$) are still slightly different with each other, which can be further captured by our proposed temporal dissimilarity mesure (${\rm TD}_{\textit{sub1-4},\textit{sub1-5}}=0.0032$ in Figure~\ref{fig:static-tempo}E).

\subsection*{Applications of temporal network comparison}
\subsubsection*{Temporal network classification}

In general, the 17 empirical temporal networks used in this work can be classified into four categories according to the contact type and venues where they were recorded (see Table~\ref{TB:1}). We firstly construct a dissimilarity matrix where each element represents the value of temporal dissimilarity of the corresponding two temporal networks (for the full matrix, see Table~S1 in \textbf{Supplementary Note 2}). We then adopt multidimensional scaling map~\cite{borg2005modern} to show the distance between them in a geometric space (Figure~\ref{fig:tempo-net_MDS}A). It shows that the four categories are spontaneously clustered into two major regions, upper right and lower left (gray shadow). In general, networks from the same category are more likely to be clustered geometrically in the two-dimensional space, except for \textit{EEU3} and \textit{ME}, two essential email networks yet now are situated in the left corner with those of schools and conferences. To obtain deep understanding, we further study the topological and temporal properties, including (i) link density; (ii) average node degree in the corresponding static networks; (iii) coefficient of variation of the node lifespan ($CV(\Delta t)$, defined as the $\delta_{\Delta t}/\mu_{\Delta t}$ where $\Delta t$ is the time difference of a node's first and last occurrence, $\delta$ and $\mu$ are respectively the standard deviation and mean; 
(iv) fraction of reachable node pairs via temporal paths ($\tilde{R} = \frac{R}{N(N-1)}$, where $R$ is the number of reachable node pairs and $N$ is the number of nodes); and (v) fraction of nodes $\tilde{N}_p$ in an evolutionary time window $[0, p*T]$, where $p \in [0,1]$. Results in Figure~\ref{fig:tempo-net_MDS}B-\ref{fig:tempo-net_MDS}E show that the networks in the same region (gray shadowed or not) in Figure~\ref{fig:tempo-net_MDS}A tend to have similar topological and temporal properties. It is observed that networks clustered in the upper region of Figure~\ref{fig:tempo-net_MDS}A show low link density, average degree, $\tilde{R}$ but high $CV(\Delta t)$, and vice verse for lower region (gray shadowed) in Figure~\ref{fig:tempo-net_MDS}A. It also clearly shows that \textit{EEU3} and \textit{ME} have similar topological and temporal properties with those of schools and conferences, which is additionally demonstrated by the evolutionary process of $\tilde{N}_p$ in Figure~\ref{fig:tempo-net_MDS}F. This suggests that those two virtually connected data may have coincident interaction patterns with the physical contact networks.

\subsubsection*{Temporal network spreadability}

\begin{figure*}[!ht]
\centering
	\includegraphics[width=18cm]{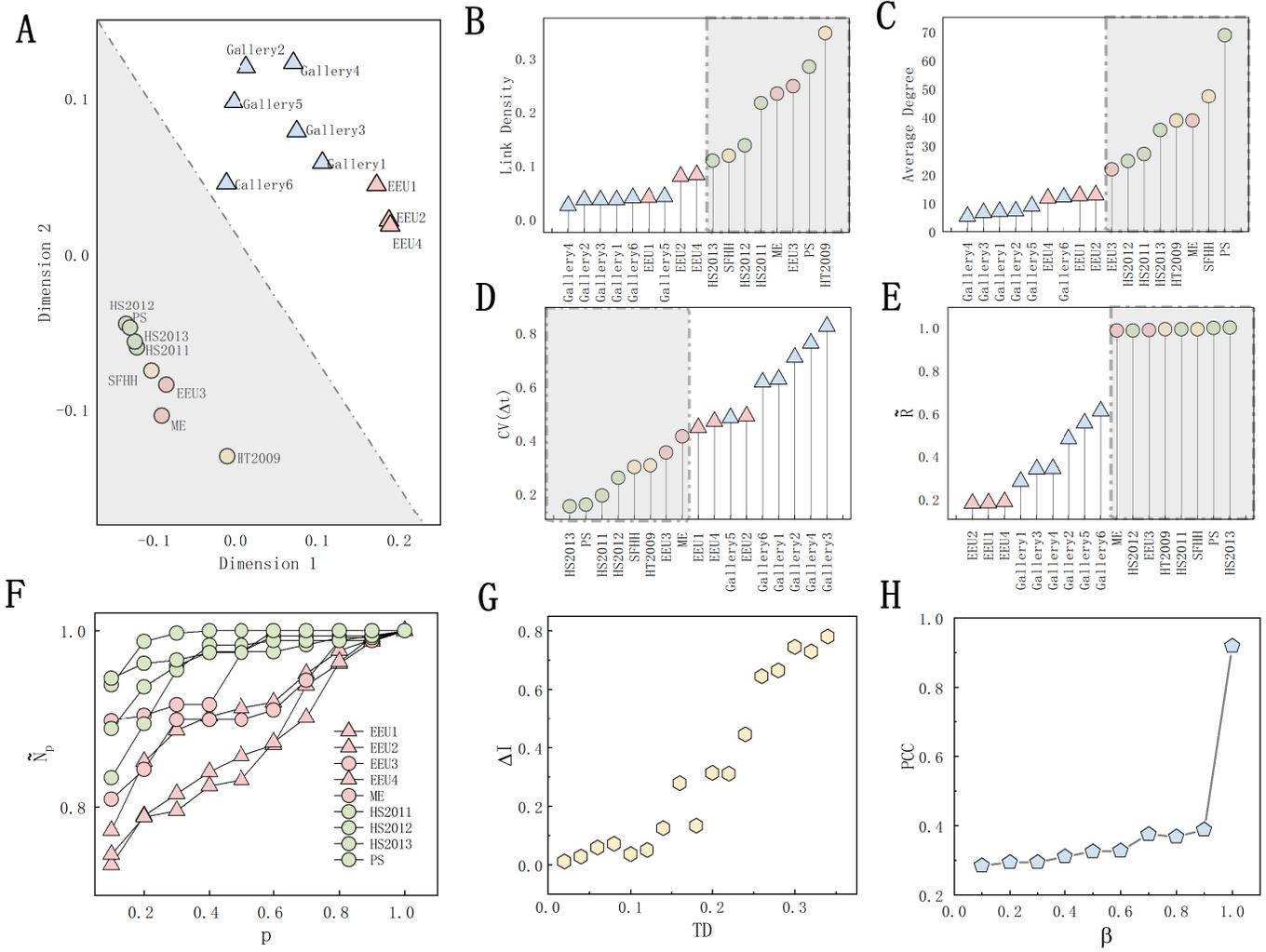}
 \caption{\textbf{Temporal network classification and spreading dynamics}. (A)
 Multidimensional scaling map of temporal dissimilarity between 17 empirical temporal networks. The colors represent different categories of networks in Table~\ref{TB:1}. (B-E) topological and temporal properties of every empirical temporal network. Gray shadows indicate the same networks at the lower left corner in (A). (B-C) Link density (B) and average node degree (C) of the regated static networks. (D-E) Coefficient of variation of node lifespan (D) and fraction of reachable node pairs via temporal paths (E) for each temporal network. (F) The fraction of nodes as the function of network evolution ($p$). (G) Spreadability differences as the function of temporal dissimilarity between 17 temporal networks. (H) Pearson correlation between spreadability differences and temporal dissimilarity as the function of infection probability $\beta$.}
 \label{fig:tempo-net_MDS}
\end{figure*}

Spreading dynamics is one of the most important functions of complex networks~\cite{pastor2015epidemic}. Here, we conduct the Susceptible-Infected (SI) spreading process in temporal networks~\cite{zhan2020susceptible}. Initially, an arbitrary node $i$ is randomly chosen as the infected \textit{seed} (I-state), and all the remaining nodes are set as the susceptible individuals (S-state). The subsequent spreading process shall follow the time step of the contacts. That is to say, at every time step $t\in[0, T]$, an I-state node will only infect its neighbors through the temporal contacts exactly existing at $t$ with infection probability $\beta$. The spreading process comes to the end at time step $T$. The fraction of final infected nodes is averaged by setting every node as the seed: $\langle I\rangle=\frac{1}{N}\Sigma_{i=1}^{N}I_{i}$, where $I_i$ is the final fraction of infected nodes by setting $i$ as the seed. As a consequence, the spreadability difference can be immediately obtained as $\Delta I = |\langle I_{G1}\rangle-\langle I_{G2}\rangle|$, where $\langle I_{G1}\rangle$ and $\langle I_{G1}\rangle$ are respectively the average fraction of final infected nodes of two temporal networks $G_1$ and $G_2$. Figure~\ref{fig:tempo-net_MDS}G shows the correlation between $\Delta I$ and temporal dissimilarity for every two temporal networks for $\beta=1$. The \emph{Pearson} correlation coefficient (PCC) is remarkably as high as $0.919$ (p<0.0001). That is, networks with close temporal dissimilarity tend to have similar spreadability in the time-varying interacted structure. Figure~\ref{fig:tempo-net_MDS}H further demonstrates that such correlation is becoming more and more obvious as $\beta$ increases.


\clearpage
\newpage
\section*{Conclusions and Discussion}\label{Conclusion}
In this work, we propose to use the fastest arrival distance (FAD) and spectral entropy based Jensen-Shannon divergence to characterize the temporal dissimilarity of temporal networks. 
To evaluate the proposed measure, we have performed comprehensive analyses on both synthetic and empirical temporal networks. Experimental results show that the proposed temporal dissimilarity can successfully discriminate temporal networks known to have different structures. Furthermore, its good performance in temporal network classification and spreadability discrimination suggests that the proposed measure could be a good indicator of the functional differences of various temporal networks with diverse network sizes and time scales by detecting the subtle structural distinctions.

Our temporal dissimilarity method belongs to the category of using graph distances and the Jensen-Shannon divergence for network comparison~\cite{schieber2017quantification}. This method can also be generalized to other network types, such as multi-layer and signed networks. The temporal aspect of a network enables to define the paths~\cite{wu2014path} between nodes in different ways, which could be a promising future work in characterizing the dissimilarity between nodes. We deem that other network comparison methodologies, such as network portraits~\cite{bagrow2019information}, communicability~\cite{chen2018complex}, Laplacian spectral~\cite{kondor2016multiscale}, persistent homology~\cite{sizemore2017classification}, can also be promising methods for temporal network comparison with appropriate definition, hence may boost the studies of temporal structures and functions, e.g., network topology variation identification~\cite{su2021identification}, node similarity characterization~\cite{chen2020framework}, vital node identification~\cite{kitsak2010identification}, community detection~\cite{peixoto2017modelling} and network synchronization~\cite{zhang2021designing}.

\section*{Materials and Methods}
\subsection*{Description of empirical temporal networks }
The temporal network datasets are given as follows:

\begin{description}
\item[Email-EU-core temporal networks (EEU)~\cite{paranjape2017motifs}] are email contact networks between members in a large European research institution. We have four respective networks, i.e., \textit{EEU1}, \textit{EEU2}, \textit{EEU3}, \textit{EEU4}, representing the email contacts between members of four different departments at the institution.
 \item[Manufacturing Email (ME)~\cite{michalski2011matching}] is an email contact network between individuals in a manufacturing company.
 \item[Gallery networks~\cite{van2012making}] are physical proximity networks of visitors of a science gallery during 69 days. A contact represents two visitors being within 1--1.5m during a 20 seconds interval. We restrict ourselves to the six first days of this data set---\textit{Gallery1} through \textit{Gallery6}.
 \item[High School datasets~\cite{fournet2014contact,mastrandrea2015contact}] contain proximity events between high school students in Marseilles, France. The datasets are recorded by sensors at 20 seconds intervals and contain data in the year of 2011, 2012 and 2013. Therefore, we can form three different temporal networks, i.e., \textit{High School 2011 (HS2011)}, \textit{High School 2012 (HS2012)}, \textit{High School 2013 (HS2013)}.
\item[Primary School (PS)~\cite{stehle2011high}] contains physical contacts between the children and teachers in a primary school.
\item[Hypertext 2009 (HT2009)~\cite{isella2011s}] contains physical contacts between attendees in the ACM Hypertext 2009 conference.
\item[SFHH conference data (SFHH)~\cite{genois2018can}] contains physical contacts between attendees in the 2009 SFHH conference in Nice, France.
\end{description}

\begin{table*}[htb]\scriptsize
\centering
\caption{\label{TB:1} Basic statistics of real-world temporal networks. $N$ represents the number of nodes, $C$ represents the total number of contacts (temporal edges), $T$ represents the duration of the observation time window, $M$ represents the number of links, $S$ represents the link density, and $\langle$ FAD $\rangle$ represents the average FAD. $M$ and $S$ are calculated according to the aggregated unweighted static networks $G$, which can be obtained by aggregating the contacts between each node pair. $\langle$ FAD $\rangle$ is averaged among the nodes that are reachable by fastest arrival path, regardless of unreachable pairs.}
\newcommand{\minitab}[2][1]{\begin{tabular}{#1}#2\end{tabular}}
\begin{tabular}{ccccccccc}
\hline
\multirow{2}*{Networks} & \multirow{2}*{$N$} & \multirow{2}*{$C$} & \multirow{2}*{$T$} & \multirow{2}*{$M$} & \multirow{2}*{\minitab[c]{$S$}} &\multirow{2}*{\minitab[c]{$\langle$ FAD $\rangle$}} &\multirow{2}*{\minitab[c]{Type} }\\\\\hline
EEU1& 309 & 61,046 & 35,097 & 3,031 &0.04 &5.14 &\multirow{5}*{\normalsize email}\\
EEU2& 162 & 46,772 & 32,340 & 1,772 &0.08 &4.34 &\\
EEU3& 89 & 12,216 & 8,911 & 1,506 &0.25 &6.12 &\\
EEU4& 142 & 48,141 & 26,496 & 1,375 &0.08 &4.12 &\\
ME& 167 & 82,876 & 57,791 & 3,250 & 0.23 &6.26 &\\\hline
Gallery1 & 200 & 5,943 & 1,238 & 714 & 0.04 &6.86 &\multirow{6}*{\normalsize visitor}\\
Gallery2 & 204 & 6,709 & 1,311 & 739 & 0.04 &10.16 &\\
Gallery3 & 186 & 5,691 & 1,240 & 615 & 0.04 &7.72 &\\
Gallery4 & 211 & 7,409 & 1,398 & 563 & 0.03 &9.64 &\\
Gallery5 & 215 & 7,634 & 975 & 967 & 0.04 &9.96 &\\
Gallery6 & 305 & 13,281 & 1,024 & 1,847 &0.04 &9.26 &\\\hline
HS2011 & 126 & 28,560 & 5,609 & 1,709 &0.22 &6.65 &\multirow{4}*{\normalsize school}\\
HS2012 & 180 & 45,047 & 11,273 & 2,220 &0.14 &7.66 &\\
HS2013 & 327 & 188,508 & 7,375 & 5,818 &0.11 &8.45 &\\
PS & 242 & 125,773 & 3,100 & 8,317 &0.29 &8.52 \\\hline
HT2009 & 113 &20,818 & 5,246 & 2,196 &0.35 &4.77 &\multirow{2}*{\normalsize conference}\\
SFHH &403 &70,261 &3,509 &9,565 &0.12 &6.71 &\\
\hline
\end{tabular}
\end{table*}
\subsection*{Activity driven model}
Given $N$ nodes, we assign every node $i$ with an active probability $a_i$, which is defined as the probability of creating a new contact with another node at time step $t$. We assume $a_i$ is extracted from an activity distribution $F(a)$, and the window of the network is $[0,T]$. The temporal network is then generated by the following steps:
\begin{itemize}
\item At each time step $t$, we assume a network $G^t$ having $N$ nodes and no contact.
\item Every node $i$ is activated with probability $a_i$ and connects to $m$ randomly selected nodes. All the new contacts at time step $t$ are added to $G^t$.
\end{itemize}


Following these procedures, we can obtain a temporal network $G (F(a),m)=\{G^t\}_{t=1}^T$, in which $F(a)$ controls the node activity and $m$ determines the number of contacts that every active node releases.

\subsection*{Temporal reference models}
In this work, we adopt six representative temporal reference models as follows (Table~\ref{refmodel} shows the static and temporal properties that are preserved or destroyed for each reference model):
\begin{description}
\item[The equal-weight Link-sequence shuffled model (EWLSS)] The Link time sequences are exchanged uniformly at random between links with the same weights, i.e., number of contacts.

\item[Link-sequence shuffled model (LSS)] The link time sequences are exchanged uniformly at random between randomly chosen links.

\item[Time shuffled model (TS)] For all the contacts in a temporal network, the timestamps are randomly reshuffled.

\item[Uniformly random times (URT)] The timestamp of every contact is obtained uniformly at random from the given observation time window $[0, T]$.

\item[Configuration model (CM)] A connected random static network $G'$ is generated via the configuration model based on the given degree sequence of static network aggregated from the corresponding temporal network $ G $, of which the timestamps are then randomly assigned on each link of $G'$.

\item[Random network (RN)] A connected Erd\H{o}s-R\'{e}nyi network is generated based on the given number of nodes and links of the aggregated static network of a temporal network, of which the timestamps are then randomly assigned on each link of the obtained Erd\H{o}s-R\'{e}nyi network.
\end{description}

\newcommand{\cmark}{\ding{51}}%
\newcommand{\xmark}{\ding{55}}%
\begin{table*}[htb]
\centering\large
\caption{\label{refmodel} \textbf{Properties that are preserved (\ding{51}) and destroyed (\ding{55}) in the corresponding reference models}, including three topological properties, degree distribution (DD), static configuration (SC) and weight correlation (WR), and three temporal properties, global timestamp sequence (GTS), the sorted contact sequence list on each link (LCS) and the whole contact sequence (WCS).}
\newcommand{\minitab}[2][1]{\begin{tabular}{#1}#2\end{tabular}}
\begin{tabular}{c||cccccc}
\hline
\multirow{2}*{Reference models} & \multirow{2}*{DD} & \multirow{2}*{SC} & \multirow{2}*{WR} & \multirow{2}*{GTS} & \multirow{2}*{LCS} & \multirow{2}*{WCS} \\\\\hline\hline
EWLSS & \ding{51} & \ding{51} & \ding{51} & \ding{51} &\ding{51} &\ding{55} \\
LSS& \ding{51} & \ding{51} & \ding{55} & \ding{51} & \ding{51} & \ding{55} \\
TS& \ding{51} & \ding{51} & \ding{51} & \ding{51} & \ding{55} & \ding{55} \\
URT& \ding{51} & \ding{51} & \ding{51} & \ding{55} & \ding{55} & \ding{55} \\
CM& \ding{51} & \ding{55} & \ding{55} & \ding{51} & \ding{51} & \ding{55} \\
RN & \ding{55} & \ding{55} & \ding{55} & \ding{51} & \ding{51} &\ding{55} \\
\hline
\end{tabular}
\end{table*}

\clearpage
\newpage

\bibliographystyle{elsarticle-num}
\bibliography{Tdis.bib}

\section*{Acknowledgements}

This work was supported by Zhejiang Provincial  Natural Science Foundation (Grant No. LR18A050001), the National Natural Science Foundation of China (Grant Nos. 61873080, and 61673151),  and the Major Project of The National Social Science Fund of China (Grant No. 19ZDA324). P.H. was supported by JSPS KAKENHI JP21H04595. HW would like to thank Netherlands
Organisation for Scientific Research NWO (TOP
Grant no. 612.001.802).

\section*{Author contributions statement}

All authors planed the study; All authors performed the experiments and prepared the figures. All authors analyzed the results and wrote the manuscript. All authors read and approved the final manuscript.

\section*{Additional information}
We use open data which can be downloaded on \url{http://www.sociopatterns.org} and \url{https://snap.stanford.edu/data/}. The
source code will be available from the first author based on reasonable request.


\end{document}